\documentclass[12pt]{iopart}
\usepackage{rotating}
\usepackage{graphicx}
\usepackage{latexsym}
\usepackage{amsfonts,bm}

\begin{document}

\title[Bounds on connective constants]
{Improved lower bounds on the connective constants for 
two-dimensional self-avoiding walks}

\author{Iwan Jensen}
\address{
ARC Centre of Excellence for Mathematics and Statistics of Complex Systems, \\
Department of Mathematics and Statistics, 
The University of Melbourne, Victoria 3010, Australia}

\ead{I.Jensen@ms.unimelb.edu.au}

\submitto{\JPA}


\begin{abstract}
We calculate improved lower bounds for the connective constants for self-avoiding
walks on the square, hexagonal, triangular, $(4.8^2)$, and $(3.12^2)$ lattices.
The bound is found by Kesten's method of irreducible bridges. This involves using
transfer-matrix techniques to exactly enumerate the number of bridges of a given span 
to very many steps. Upper bounds are obtained from recent exact
enumeration data for the number of self-avoiding walks and compared to current
best available upper bounds from other methods.
\end{abstract}

\maketitle

\section{Introduction}

The self-avoiding walk (SAW) on regular lattices is one of the most important 
and classic combinatorial problems in statistical mechanics \cite{MSbook}. 
An {\em $n$-step self-avoiding walk} $\bm{\omega}$ on a regular lattice is 
a sequence of {\em distinct} vertices $\omega_0, \omega_1,\ldots , \omega_n$ 
such that each vertex is a nearest neighbour of it predecessor. SAWs are
considered distinct up to translations of the starting point $\omega_0$.
The fundamental problem is the calculation (up to translation) of the number 
of SAWs, $c_n$, with $n$ steps. It is generally believed that $c_n$ grows
exponentially with power law corrections

\begin{equation}\label{eq:cn}
c_n\sim A \mu^n n^{\gamma-1},
\end{equation} 
\noindent
where $\mu$ is called the {\em connective constant}, $\gamma$ is
a critical exponent and $A$ a critical amplitude. Hammersley and Morton \cite{HM54}
were the first to prove the existence of the limit

\begin{equation}\label{eq:mulim}
\mu = \lim_{n\to \infty} c_n^{1/n}
\end{equation}
\noindent
The exact value of $\mu$ is known only on the hexagonal lattice, where 
Nienhuis \cite{Nienhuis82a,Nienhuis84a} showed, using non-rigorous methods, that
$\mu_{\rm hex}=\sqrt{2+\sqrt{2}}$, and on the $(3.12^2)$ lattice, where Jensen and 
Guttmann \cite{JG98} found an exact and rigorous connection between the connective 
constant $\mu_{(3.12^2)}$  and the connective constant for the hexagonal
lattice $\mu_{\rm hex}=\mu_{(3.12^2)}^{3}/(\mu_{(3.12^2)}+1)$. 
On the square lattice it has been observed \cite{CEG93} that $\mu_{\rm sq}$ is 
indistinguishable from the reciprocal of the unique positive root $x_c$ 
of the simple polynomial  $581x^4 + 7x^2 -13=0$, and while this `conjecture' 
has stood the test of time it remains a purely numerical observation.

Since finding the exact value of $\mu$ (let alone proving such results rigorously)
is extremely difficult much effort has been devoted to more general methods
for proving rigorous {\em bounds} on the connective constant. Brief overviews
of some of the methods used can be found in \cite{MSbook,AJG83a}. A systematic
procedure for improving the lower bounds can be devised from a method due to
Kesten \cite{Kesten63a}. It was used by Guttmann \cite{AJG83a} to improve the
lower bounds for the connective constant on the square and simple cubic lattices 
and more recently by Alm and Parviainen \cite{Alm04} to obtain improved lower
bounds on the connective constant for the hexagonal lattice. In this paper
we further refine these bounds and extend the work to the triangular, kagom\'e and
$(4.8^2)$ lattices. 

Finally, we use recent exact enumeration data for $c_n$ to obtain upper bounds
for the connective constant on the square, hexagonal and triangular lattices.
These bounds are then compared to better upper bounds obtained from other methods.

\section{Lower bounds}

Lower bounds for the connective constant can be found using a method
due to Kesten \cite{Kesten63a}. The method utilises the fundamental result 
that certain restricted classes of self-avoiding walks have the same connective
constant as the unrestricted problem. Particularly useful for our purposes
is the class of walks known as {\em bridges}. Let $x_j$ denote the
$x$-coordinate of $\omega_j$, then a bridge is a self-avoiding walk
such that $x_0 < x_j \leq x_n$ for all $j>0$. We use $b_n$ to denote
the number of $n$-step bridges, and note  Kesten showed that
$b_n^{1/n}$ converges to $\mu$ as $n\to \infty$. Clearly concatenating two bridges
of length $n$ and $m$ gives a bridge of length $n+m$ (we place the origin of
the second walk on top of the end-point of the first walk). This means
that any bridge can be decomposed into {\em irreducible bridges}, i.e.,
bridges which cannot be decomposed further, and we use $a_n$ to
denote the number of $n$-step irreducible bridges. It is now easy
to see that the generating function $B(x)$ for bridges is simply related 
to the generating function for irreducible bridges $A(x)$

$$B(x)=\frac{1}{1-A(x)}.$$
\noindent
It then follows that $1/\mu$ is the solution to $A(x)=1$. This relation
also allows us to obtain lower bounds for $\mu$. This relies on
the observation that, if $0 \leq \tilde{a}_n \leq a_n$, for $n\geq 2$,
then with $x_c$ being the solution to 

\begin{equation}\label{eq:lowergen}
\sum_{n=1}^{\infty} \tilde{a}_nx^n = 1
\end{equation}
\noindent
$1/x_c$ is a lower bound on $\mu$. In particular we can set $\tilde{a}_n=0$
for $n>N$ and thus truncate the series. 

It is not easy to calculate the number of irreducible bridges directly.
Thankfully they can easily be obtained from the number of bridges.
Following Alm and Parviainen \cite{Alm04} we consider the number of bridges 
$b_{n,l}$ and irreducible bridges $a_{n,l}$ of length $n$ and span $l$, 
that is bridges with $x_0=0$ and $x_n=l>0$, with associated generating
functions $B_l(x)$ and $A_l(x)$. Obviously $\sum_{l=1}^{\infty}a_{n,l} = a_n$,
so if we truncate at some maximum span $L$ and maximum walk length $N$ then
the reciprocal of the solution to

\begin{equation}\label{eq:lower1}
\sum_{n=1}^N \sum_{l=1}^L a_{n,l} x^n =1
\end{equation}
\noindent
is a lower bound on $\mu$. 

Since a bridge is either irreducible or the concatenation of a bridge
with an irreducible bridge we get

$$B_l(x) = A_l(x) + \sum_{k=1}^{l-1} A_{l-k}(x)B_k(x)$$
\noindent
and thus 

$$A_l(x) = B_l(x)-\sum_{k=1}^{l-1} A_{l-k}(x)B_k(x),$$
\noindent
which allows to obtain all generating functions $A_l(x)$ recursively from 
$B_l(x)$ for $1 \leq l \leq L$. 

In this paper we also examine a second way of obtaining lower bounds.
We again use irreducible bridges, but rather than using smallish $L$ 
and very large $N$ we calculate the exact series for irreducible bridges
to order $N$ (much lower than before) and use this truncated series
to obtain a lower bound from the reciprocal of the solution to

\begin{equation}\label{eq:lower2}
\sum_{n=1}^N  a_{n} x^n =1,
\end{equation}
\noindent
that is in Eq.~(\ref{eq:lowergen}) we set $\tilde{a}_n=a_n$ for $n\leq N$ 
and $\tilde{a}_n=0$ for $n>N$.

\subsection{Enumeration of self-avoiding bridges \label{sec:flm}}

The number of self-avoiding bridges $b_{n,l}$ can easily be counted using the
Transfer-Matrix (TM) methods we have developed for the unrestricted problems 
\cite{IJ04a,IJ04d,IJ04f}, which are devised to count the number of walks in
a finite $l\times w$ rectangular sub-section of the underlying lattice. Here we 
shall only briefly outline the changes required to enumerate bridges. 
The most efficient implementation of the 
TM algorithm generally involves bisecting the rectangle with a boundary line 
and moving the boundary in such a way as to build up the lattice cell by cell. 
The sum over all contributing graphs is calculated as the boundary is moved 
through the lattice. For each configuration of occupied or empty edges 
along the intersection we maintain a generating function for partial walks 
cutting the intersection in that particular pattern. If we draw a SAW and 
then cut it by a line we observe that the partial SAW to the left of this 
line consists of a number of loops connecting two edges in the intersection, and 
at most two pieces connected to only one edge (these are the pieces from the 
end-points of the SAW). The computational complexity of 
the algorithm is essentially determined by the number of such configurations.
So we must make the intersection as short as possible. Since we are looking to
fix $l$ and make $N$ large it follows that $w$ will be large as well (in fact
proportional to $N-l$).
So the boundary line must intersect the rectangle along the `bridging' axis, e.g., 
along up to $l+2$ edges. It is quite easy to demonstrate \cite{CEG93,IJ04a} 
that the number of configurations grows like $3^l$ in the square lattice case. 
So the required CPU time will grow roughly as $(w+l)^23^l=N^2 3^l$, since there 
are $(w+l)=N$ updates and terms in the generating functions. Memory requirement 
will grow as $N 3^l$. We note in passing that while the TM method can be used
to study higher-dimensional lattices it quickly becomes inefficient because
the boundary would be $(d-1)$ dimensional and the number of edges in the
intersection would grow ever more rapidly.

\begin{figure}
\begin{center}
\includegraphics{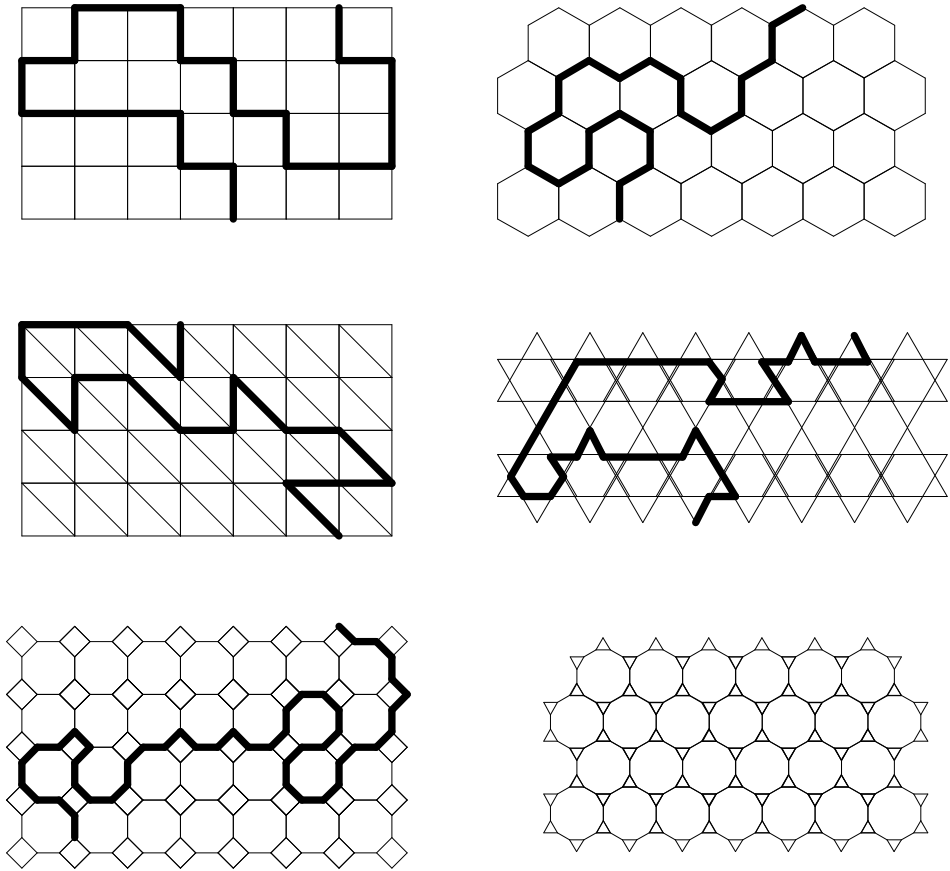}
\end{center}
\caption{\label{fig:bridges}
Examples of bridges on the square, hexagonal, triangular, kagom\'e and $(4.8^2)$ lattices.
In addition we also show a section of the $(3.12^2)$ lattice.}
\end{figure}

In order to implement the first method for finding lower bounds, 
see Eq.~(\ref{eq:lower1}), we count the number of bridges spanning rectangles of 
size $l\times w$, that is bridges starting at the bottom border and terminating 
at the top border. In addition the walks must also touch the left border of the 
rectangle (this takes care of the translational invariance) as illustrated in 
Figure~\ref{fig:bridges}. In all case bridges must terminate at a top most vertex 
in the top most row. Note in particular the implication of this restriction on the 
hexagonal, kagom\'e, and $(4.8^2)$ lattices. For the hexagonal case this means that 
all bridges are of even length.

For the square lattice we calculated the number of bridges to $L=15$ and
$N=250$, for the hexagonal lattice to $L=15$ and $N=500$, 
for the triangular lattice to $L=12$ and $N=150$, 
for the kagom\'e lattice to $L=10$ and $N=300$,and
for the $(4.8^2)$ lattice to $L=12$ and $N=500$.
Because of the exact connection between the connective constants $\mu_{(3.12^2)}$ and
$\mu_{\rm hex}$, $\mu_{\rm hex}=\mu_{(3.12^2)}^{3}/(\mu_{(3.12^2)}+1)$, 
any bounds for the hexagonal lattice yields corresponding bounds for the 
$(3.12^2)$ lattice. So we don't actually count bridges on the $(3.12^2)$ lattice
and have thus not shown an example of one in Figure~\ref{fig:bridges}.

The integer coefficients occurring in the series expansions become very large.  
The calculations were therefore performed using modular arithmetic \cite{KnuthACPv2}. 
This involves performing the calculation modulo various integers $p_i$ and then 
reconstructing the full integer coefficients at the end. The $p_i$ are called 
moduli and must be chosen so they are mutually prime. The Chinese remainder theorem 
ensures that any integer has a unique representation in terms of residues. If the 
largest value occurring in the final expansion is $m$, then we have to use a number 
of moduli $k$ such that $p_1p_2\cdots p_k > m$. We used moduli which are prime numbers 
of the form $p_k=2^{30}-r_k$.  

Naturally the calculation for each $l$ and moduli are completely independent.
It is evident from the exponential growth in the computational complexity
most of the CPU time is spent on the largest value of $L$, where up to
16 moduli were required to represent the coefficients. Typically 
each run (at the maximal span $L$) required up to 24 CPU hours on
a 1GHz Alpha processor and could use up to 2.5Gb of memory. In all we
used about 3000 CPU hours on the calculations. Our method is much more
efficient than that used by Alm and Parviainen \cite{Alm04}, who report using 
more than 20000 CPU hours calculating the number of bridges on the hexagonal 
lattice with $L=10$ and $N=58$. A similar calculation using our method takes
no more than a couple of minutes!

The second method for finding lower bounds, see Eq.~(\ref{eq:lower2}), uses the exact
data for the number of irreducible bridges up to length $N$. Again the first step
is the calculation of the relevant data for bridges. We illustrate the method in the
square lattice case. An irreducible bridge of width $L$ has length at least $3L$.
This is because each row (apart from the bottom most) must have more than one 
occupied edge (otherwise we could cut the walk into two bridges) and the walk
must thus go up, come down, and go up again. We also have the first step and
at least two horizontal steps for a grand total of at least $3L$ steps. 
So if we require the number of irreducible bridges to order $N$ we must
count the number of bridges with span up to $L=N/3$. That is we have to  count
the number of bridges on rectangles of size $w\times l$, where $1 \leq l \leq L=N/3$
and $1 \leq w \leq N-l$. 
Note that this calculation gives the number of bridges correctly only to order $L$.
However, by first extracting the series for $A(x)$ we can also get $B(x)=1/(1-A(x))$
correct to order $3L$. The efficient calculation of the bridge generating function is 
in many aspects more complicated and time consuming than for the first method.
Details of the properties of the bridge generating function will appear in a
separate paper. Suffice to say that we have obtained generating functions to
order 72 on the square lattice and 122 on the hexagonal lattice.

The series for the problems studied in this paper can be obtained by request 
from the author or at http://www.ms.unimelb.edu.au/\~{ }iwan/ by following the relevant 
links.

\subsection{Results}

\begin{table}
\caption{\label{tab:hexlower} Lower bounds for the connective constant
for the hexagonal lattice. The supposed exact value is
$\mu=\sqrt{2+\sqrt{2}}=1.847759065\ldots$.
}
\begin{indented}
\item[]\begin{tabular}{llllll} \br
L     & $N=100$   & $N=200$   & $N=300$   & $N=400$   & $N=500$  \\ 
\mr
2    &   1.787760 &   1.787760 &   1.787760 &   1.787760 &   1.787760 \\ 
3    &   1.808678 &   1.808678 &   1.808678 &   1.808678 &   1.808678 \\ 
4    &   1.819369 &   1.819370 &   1.819370 &   1.819370 &   1.819370 \\ 
5    &   1.825750 &   1.825767 &   1.825767 &   1.825767 &   1.825767 \\ 
6    &   1.829869 &   1.829984 &   1.829984 &   1.829984 &   1.829984 \\ 
7    &   1.832546 &   1.832950 &   1.832951 &   1.832951 &   1.832951 \\ 
8    &   1.834182 &   1.835137 &   1.835140 &   1.835140 &   1.835140 \\ 
9    &   1.835061 &   1.836802 &   1.836814 &   1.836814 &   1.836814 \\ 
10   &   1.835448 &   1.838094 &   1.838132 &   1.838132 &   1.838132 \\ 
11   &   1.835574 &   1.839101 &   1.839191 &   1.839193 &   1.839193 \\ 
12   &   1.835602 &   1.839878 &   1.840058 &   1.840063 &   1.840063 \\ 
13   &   1.835606 &   1.840464 &   1.840775 &   1.840788 &   1.840789 \\ 
14   &   1.835606 &   1.840890 &   1.841372 &   1.841400 &   1.841402 \\ 
15   &   1.835606 &   1.841184 &   1.841868 &   1.841921 &   1.841925 \\ 
\br
\end{tabular}
\end{indented}
\end{table}
 
Lower bounds are obtained by forming the polynomials of Eq.~(\ref{eq:lower1}).
It is thus possible to obtain ever improved lower bounds by increasing
$N$ and $L$. In Table~\ref{tab:hexlower} we use the hexagonal lattice
data to illustrate the method. Note that for this problem the minimal
number of steps for an irreducible bridge of span $L$ is $6L-2$.
From this data we observe first of all, that for $N=100$ little is gained 
by going beyond span $L=12$. This is somewhat surprising since $A_{13}(x)$ 
contributes already at order 76, but obviously the influence of these terms
is almost negligible. Likewise with fixed $L$ and increasing $N$ it is a
case of rapidly diminishing returns. If we are interested in optimising
the procedure, that is, getting a decent bound, but with as little wasted
effort possible, it appears that for given $L$ we should choose $N$ larger than
twice the order of the first non-zero contribution to $A_L (x)$ (otherwise the 
calculation of $A_L (x)$ is largely wasted) but not much larger than four times 
this order. Similar considerations apply to the other problems as well though
the optimal cut-off varies from problem to problem.

Here we briefly summarise our results for the lower bounds. For the hexagonal lattice 
we find the lower bound, $1.841925 < \mu_{\rm hex}$, which is less than 0.32\% lower than the 
exact value $\mu_{\rm hex} = \sqrt{2+\sqrt{2}}=1.847759065\ldots$.
The previous best lower bound was  $1.833009 < \mu_{\rm hex}$ \cite{Alm04}.
For the square lattice  we obtain the lower bound, $2.625622 < \mu_{\rm sq}$, 
which is within 0.48\% of the best estimate for the connective constant 
$\mu_{\rm sq}=2.63815853031(3)$ \cite{IJ03a}. This should be compared to
the previous bound $2.62006 < \mu_{\rm sq}$ \cite{CG93}.
For the triangular lattice the lower bound is, $4.118935 < \mu_{\rm tri}$,  
within 0.77\% of the best estimate $\mu_{\rm tri}=4.150797226(26)$ \cite{IJ04d},
whereas the previous best bound was $4.03333 < \mu_{\rm tri}$ \cite{Almbound}.
The Kagom\'e  lattice lower bound is, $2.548497<\mu_{\rm kag}$, within 0.48\% of 
the estimate $\mu_{\rm kag}=2.560576765(10)$ (based on our unpublished enumerations 
of self-avoiding polygons), while the previous best bound was  
$2.50967<\mu_{\rm kag}$ \cite{Almbound}.
For the $(4.8^2)$ lattice we found the lower bound,  $1.804596<\mu_{(4.8^2)}$, which
is just 0.24\% lower that the estimate $\mu_{(4.8^2)}=1.80883001(6)$ \cite{JG98},
which improves on the bound $1.78564<\mu_{(4.8^2)}$ \cite{Almbound}. 
Finally, for the $(3.12^2)$ lattice we get the lower bound, $1.708553< \mu_{(3.12^2)}$, 
which is just 0.15\% from the exact value $\mu_{(3.12^2)}=1.711041296\ldots$ \cite{JG98},
and again improves on the previous bound $1.705263< \mu_{(3.12^2)}$ \cite{Alm04}.

\begin{table}
\caption{\label{tab:loweralt} Lower bounds for the connective constant
for the hexagonal and square lattices.
}
\begin{indented}
\item[]\begin{tabular}{llll} \br
\multicolumn{2}{c}{Hexagonal} & \multicolumn{2}{c}{Square} \\ \mr
$N$    & Bound   & $N$   & Bound   \\ 
\mr
32   &   1.812833  & 30  &  2.583704 \\
38   &   1.817977  & 33  &  2.588448 \\
44   &   1.821786  & 36  &  2.592419 \\
50   &   1.824722  & 39  &  2.595794 \\
56   &   1.827055  & 42  &  2.598698 \\
62   &   1.828955  & 45  &  2.601224 \\
68   &   1.830532  & 48  &  2.603442 \\
74   &   1.831863  & 51  &  2.605405 \\ 
80   &   1.833002  & 54  &  2.607155 \\ 
86   &   1.833987  & 57  &  2.608726 \\
92   &   1.834847  & 60  &  2.610143 \\
98   &   1.835606  & 63  &  2.611428 \\
104  &   1.836279  & 66  &  2.612599 \\
110  &   1.836882  & 69  &  2.613671 \\
116  &   1.837424  & 72  &  2.614656 \\
\br
\end{tabular}
\end{indented}
\end{table} 

\begin{figure}
\begin{center}
\includegraphics[scale=0.5]{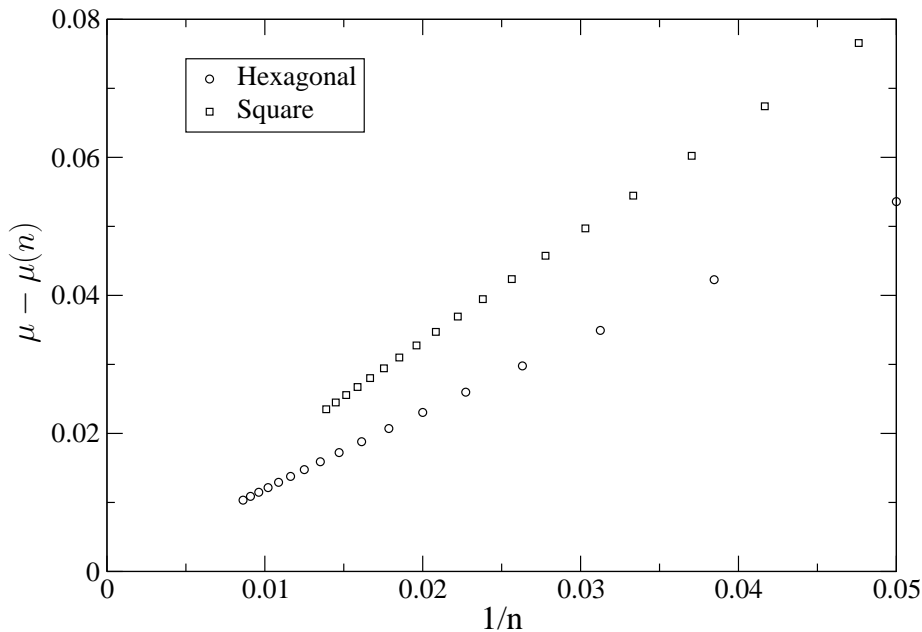}
\end{center}
\caption{\label{fig:bounds}
The difference between the connective constant $\mu$ and the lower bound $\mu(n)$ 
vs. $1/n$. Data is plotted for the hexagonal and square lattices.}
\end{figure}

As stated earlier we also used a second approach to obtain lower bounds for the 
connective constant. This entails the calculation of an exact series expansion for 
the generating function for irreducible bridges up to some maximal order $N$ 
(this was also the method employed by Guttmann \cite{AJG83a,AJG83b}). Lower bounds 
are then obtained from the truncated series in Eq.~(\ref{eq:lower2}). Obviously we could 
truncate at any order $n<N$ and obtain a sequence of lower bounds $\mu(n)$. 
In Table~\ref{tab:loweralt} we have listed the lower bounds obtained from this method 
for the hexagonal and square lattice cases. Clearly this method is inferior to
the previous one (the bounds are not as good) particularly considering that
the computational effort is significantly greater. However, this approach allows us to 
study the convergence of the lower bounds $\mu(n)$ to the connective constant as a 
function of the truncation order $n$. We find that $\mu-\mu(n) \sim a/n$, 
this behaviour can be seen directly in Fig.~\ref{fig:bounds} where we have plotted
$\mu-\mu(n)$ vs. $1/n$. We also formed the generating function 
$D(x)=\sum_n d_n x^n$, where $d_n=\mu-\mu(n)$, analysed this using differential
approximants and found a logarithmic singularity at $x_c =1$, as expected if 
$d_n \sim a/n$.

\section{Upper bounds}

The best current method for obtaining upper bounds is due to Alm \cite{Alm93}
and it essentially requires one to enumerate the number of walks according to
length $n$ and a specified head and tail each of length $m$.
More precisely Alm showed that

\begin{equation}\label{eq:Alm}
\mu \leq (\lambda(G(m,n)))^{1/(n-m)},
\end{equation}
\noindent
where $\lambda$ is the largest eigenvalue of the matrix $G(m,n)$. The entries $g_{ij}$ 
of this matrix are equal to the number of $n$-step self-avoiding walks starting with 
a walk $\bm \omega_i$ and ending with a translation of a walk $\bm \omega_j$.
Each walk $\bm \omega_i$, $i=1,\ldots, K_m$, is one of the $K_m$ possible
$m$-step self-avoiding walks (up to all possible symmetries). While this
method can yield quite sharp upper bounds (within 1.1\% for the hexagonal lattice
\cite{Alm04}) it is unfortunately not suited for a transfer-matrix enumeration.

However, we have recently obtained greatly extended series for the number of
SAWs on the square, hexagonal and triangular lattices \cite{IJ04a,IJ04d,IJ04f}.
This allows us to use a special case of Alm's work \cite{Alm93} which states
that if $K_m=1$ then 
 
\begin{equation}\label{eq:Alm1}
\mu \leq \mu_m(n) = (c_n/c_m)^{1/(n-m)}.
\end{equation}
\noindent
On all lattices $K_1=1$ so $\mu  \leq \mu_1(n)=(c_n/c_1)^{1/(n-1)}$ as proven earlier
by Ahlberg and Janson \cite{Ahlberg81} while $K_2=1$ for the hexagonal
and $L$ lattices yielding sharper bounds $\mu \leq \mu_2(n)=(c_n/c_2)^{1/(n-2)}$ as
proven by Guttmann \cite{AJG83b}. Ahlberg and Janson also proved that
an upper bound $\mu_a(n)$ can be obtained from the positive root of the equation

\begin{equation}\label{eq:Ahlberg}
zx^{n-1}= [c_n-(z-2)c_{n-1}]x+(z-2)[(z-1)c_{n-1}-c_n],
\end{equation}
\noindent
where $z=c_1$ is the coordination number of the lattice. An upper bound
is then found as $\min (\mu_m(n),\mu_a(n))$. 

For the square lattice we have $c_{71}=4190893020903935054619120005916$ and
$c_{70}=1580784678250571882017480243636$, which gives us the upper bounds
$\mu_1=2.684484$ and $\mu_a=2.681360$. These bounds are sharper than the one
obtained by Alm \cite{Alm93}, $\mu_{\rm sq} < 2.695759$, using $n=24$ and $m=8$.
An improved upper bound, $\mu_{\rm sq} < 2.679193$, has been obtained by
P\"onitz and Tillmann \cite{PT00}, by counting walks with finite memory. 

For the triangular lattice we have $c_{40}=22610911672575426510653226$ and
$c_{39}=5401678666643658402327390$, which gives us the upper bounds
$\mu_1=4.267349$ and $\mu_a=4.263713$.  These bounds are sharper than the one
obtained by Alm \cite{Alm93}, $\mu_{\rm tri} < 4.277799$, using $n=16$ and $m=6$ 
(Alm has since improved this to $\mu_{\rm tri} < 4.25152$ \cite{Almbound}).

For the hexagonal lattice we have $c_{100}=2585241775338665938539885252$ and
$c_{99}=1394474897269109512317080364$, which gives us the upper bounds
$\mu_1=1.871004$, $\mu_2=1.869731$ and $\mu_a=1.869836$. This should be
compared with the sharper bound $\mu_{\rm hex} \leq 1.868832$ obtained in \cite{Alm04}
using the method of \cite{Alm93} with $n=45$ and $m=17$. 

It is clear that sharper upper bounds can be obtained for square and
triangular lattices by Alm's method if carried out to higher values of $n$ and
$m$. Judging from the computational resources (928 CPU hours) required to
obtain the bounds in \cite{Alm04} this should not be a very demanding
calculation (compare this to the 25000 CPU hours used for the enumeration
of the hexagonal lattice SAWs).

\section{Summary}

We have used Kesten's method of irreducible bridges to obtain improved lower bounds
on the connective constant for self-avoiding walks on several planar lattices.
The number of irreducible bridges is obtained by enumerating exactly the number
of bridges using transfer-matrix techniques. In one approach we calculate the
number of bridges of limited span but to great lengths while in a second approach
we obtain an exact series expansion for the number of irreducible bridges.
The first approach turns out to yield the sharpest lower bounds. The second
approach allows us to study the convergence of the lower bounds $\mu(n)$ to the 
connective constant as a function of the truncation order $n$. We find that
the limit is approached linearly in $1/n$. In addition we use recent exact
data for the number of SAWs $c_n$ to obtain some upper bounds on the
connective constant. The upper bounds are generally much poorer than the lower bounds
and also worse than those already obtained by other methods.
We have summarised the results in Table~\ref{tab:sum}.

\begin{table}
\begin{indented}
\item[]\caption{\label{tab:sum} Summary of results for the connective constant $\mu$ with
the current best lower and upper bounds}
\begin{tabular}{llll} \br
Lattice     & Lower Bound   & $\mu$ & Upper Bound \\ 
\mr
Square     & 2.625622 & $2.63815853031(3)$ \cite{IJ03a}        & 2.679193 \cite{PT00} \\
Hexagonal  & 1.841925 & $1.847759065\ldots$ \cite{Nienhuis82a} & 1.868832 \cite{Alm04} \\
Triangular & 4.118935 & $4.150797226(26)$ \cite{IJ04d}         & 4.25152 \cite{Almbound} \\
Kagom\'e   & 2.548497 & $2.560576765(10)$                      & 2.590301 \cite{GPR04} \\
$(4.8^2)$  & 1.804596 & $1.80883001(6)$ \cite{JG98}            & 1.82926 \cite{Almbound} \\
$(3.12^2)$ & 1.708553 & $1.711041296\ldots$ \cite{JG98}        & 1.719254 \cite{Alm04} \\ 
\br
\end{tabular}
\end{indented}
\end{table}

\section{Acknowledgments}

The calculations presented in this paper were performed on the facilities of the
Australian Partnership for Advanced Computing (APAC). We gratefully acknowledge 
financial support from the Australian Research Council.

\section*{References}


\end{document}